\documentclass[a4paper,11pt]{article}
\usepackage{graphicx,amssymb,amstext,amsmath,mathabx}
\usepackage{color, yfonts, tcolorbox}
\usepackage{floatflt}

\usepackage{float}
\usepackage{esint}
\usepackage{subcaption}
\usepackage{cite}
\usepackage{mathtools}
\usepackage{commath}
\usepackage{array}
\usepackage{boldline}
\usepackage{multirow}
\usepackage{arydshln}
\usepackage{bigdelim}
\usepackage{hyperref}
\definecolor{cbl}{rgb}{0,0,1}

\newcommand{\bc}{\begin{center}}
\newcommand{\ec}{\end{center}}
\def\ba#1{\begin{array}{#1}\displaystyle}
\newcommand{\ea}{\end{array}}

\newcommand{\beq}{\begin{equation}}
\newcommand{\eeq}{\end{equation}}
\newcommand{\beqa}{\begin{eqnarray}}
\newcommand{\eeqa}{\end{eqnarray}}

\newcommand{\bi}{\begin{itemize}}
\newcommand{\ei}{\end{itemize}}

\begin{document}
\begin{titlepage}
\vspace{0.2cm}
\begin{center}

{\large{\bf{Y-Systems for Generalised Gibbs Ensembles\\ in Integrable Quantum Field Theory}}}

\vspace{0.8cm} 
{\large \text{Olalla A. Castro-Alvaredo}}

\vspace{0.8cm}
{\small  Department of Mathematics, City, University of London, 10 Northampton Square EC1V 0HB, UK\\
}
\end{center}

\medskip

The thermodynamic Bethe ansatz approach to the study of integrable quantum field theories was introduced in the early 90s. Since then it has been known that the thermodynamic Bethe ansatz equations can be recast in the form of  $Y$-systems. These $Y$-systems have a number of interesting properties, notably in the high-temperature limit their solutions are constants from which the central charge of the ultraviolet fixed point can be obtained and they are typically periodic functions, with period proportional to the dimension of the perturbing field. In this letter we discuss the derivation of $Y$-systems when the standard thermodynamic Bethe ansatz equations are replaced by generalised versions, describing generalised Gibbs ensembles. We shown that for many integrable quantum field theories, there is a large class of distinct generalised Gibbs ensembles which share the same $Y$-system. 

\medskip
\medskip

\vfill

\noindent o.castro-alvaredo@city.ac.uk

\hfill \today

\end{titlepage}
\section{Introduction}
In the context of the thermodynamic Bethe ansatz (TBA) treatment of integrable quantum field theories (IQFTs) \cite{tba1,elastic,tba2}, $Y$-systems were first introduced and studied in \cite{Z,KP,KN,RTV}.  More precisely, it was realised that the thermodynamic Bethe ansatz equations for functions $\varepsilon_a(\theta)$ commonly known as pseudoenergies ($a$ labels the particle type and $\theta$ the rapidity) can be equivalently written as systems of equations for functions $Y_a(\theta)=e^{\varepsilon_a(\theta)}$ known as $Y$-systems. The transformation from TBA-equations (which are non-linear integral equations) to $Y$-systems (which are sets of coupled algebraic equations) is deceptively simple: one Fourier-transforms the former, rearranges terms in the resulting equations and then inverse Fourier-transforms. In the process the convolution integrals typical of TBA equations are eliminated. $Y$-systems  can be constructed for discrete and lattice theories as well as for IQFTs with non-diagonal scattering. Here, however we will restrict ourselves to relativistic IQFTs, focusing on the family of minimal Toda field theories, even if our observations apply much more generally.
\medskip 

Our aim is to consider the kind of generalised TBA-equations discussed in \cite{Caux} and already implicit in \cite{FM}.  Let us start however by considering the standard equations, as proposed by Zamolodchikov \cite{tba1}. The properties of a generic  1+1D IQFT with diagonal scattering at thermal equilibrium are described by the equations
\beq
\varepsilon_a(\theta)=m_a\beta \cosh\theta-\sum_{b=1}^r (\varphi_{ab}\star L_b)(\theta)\,, \qquad a=1,\ldots, r\,,
\label{tba}
\eeq 
where $a$ specifies the particle type, $m_a$ is the mass of a particle of type $a$, $r$ is the number of particles in the spectrum, and $\beta=1/T$ is the inverse temperature. The symbol $\star$ denotes the convolution
\beq 
(a\star b)(\theta):=\frac{1}{2\pi}\int_{-\infty}^\infty d\theta' a(\theta-\theta') b(\theta')\,,
\eeq 
and the functions $L_a(\theta)$ are given by $L_a(\theta):=\log(1+e^{-\varepsilon_a(\theta)})$ for theories with fermionic statistics. The information about scattering is embedded into the kernels (or scattering phases) $\varphi_{ab}(\theta)$ which are the logarithmic derivatives of the corresponding two-particle scattering matrices $S_{ab}(\theta)$
\beq 
\varphi_{ab}(\theta):=-i \frac{d}{d\theta} \log S_{ab}(\theta)\,.
\eeq 
In recent years, the study of out-of-equilibrium many-body quantum systems has led to the realisation that IQFTs dynamically evolve towards equilibria which are described by generalised Gibbs ensembles (GGEs) \cite{rigol}. At the level of the TBA-equations, GGEs are described by generalised versions of (\ref{tba}) where the driving term $m_a \beta \cosh\theta$ is replaced by a more generic function
\beq 
\nu_a(\theta)=\sum_{i=1}^N  \lambda_a^i h_a^{i}(\theta;s_i)\,,
\eeq
where $h_a^i(\theta;s)$ are one-particle eigenvalues of local or quasi-local conserved charges $Q_a^{s}$ of spin $s$ and $\lambda_a^i$ are parameters that play the role of generalised inverse temperatures. Typically, in relativistic 1+1D quantum field theory we have that $h_a^0(\theta;0)=1$, $h_a^1(\theta;1)=m_a \cosh\theta$ and $h_a^2(\theta;1)=m_a \sinh\theta$ corresponding to particle number,  energy and momentum, respectively. In general, spin-$s$ quantities have one-particle eigenvalues which are combinations of the functions $e^{\pm s\theta}$. 

Starting with the generalised TBA-equations for a GGE in IQFT it is natural to ask what the corresponding $Y$-system would be. This is a question that has not been addressed yet in the context of IQFT, mainly because most work on GGEs and their associated Bethe ansatz description so far has focussed on quantum spin chains.  Indeed, for the latter there have been several works where $Y$-systems are discussed  \cite{IQC,LBE,PVW,LEPB1,LEPB2}.  The aim of this paper is to highlight the special features of the $Y$-systems associated to a certain class of GGEs. 

\section{A Simple Example}
One of the simplest examples where we can illustrate the construction of a $Y$-system and its generalisation to GGEs is the sinh-Gordon model at the self-dual point \cite{sinh1,sinh2,sinh3,sinh4} (see \cite{FMS} for a more recent introduction to the model's scattering theory). For our purposes, all we need to know about the model is that it is a theory of a single particle (so we can drop the indices in the kernel), and that the kernel  is given by $\varphi(\theta)=2 \,{\rm sech}\theta$. This type of kernel, which is strongly peaked around $\theta=0$ and exponentially decaying for large $\theta$ is typical of IQFTs with diagonal scattering. The $Y$-system is then constructed as follows: first, the TBA equation is Fourier-transformed to
\beq
\tilde{\xi}(k)+\frac{1}{2\pi}\tilde{\varphi}(k) \tilde{L}(k)=0\,,
\eeq
where the ``tilded" functions are all Fourier transforms and $\xi(\theta):=\varepsilon(\theta)-m\beta \cosh\theta$. Here we exploited the fact that the Fourier transform of a convolution is the product of the Fourier transforms (up a normalisation constant depending on the normalisation of the Fourier transform). For the kernel above we have that $\tilde{\varphi}(k)=2\pi \,{\rm sech} \frac{\pi k}{2}$ 
and so the equation can be rewritten as
\beq
\cosh \frac{\pi k}{2} \tilde{\xi}(k)+ \tilde{L}(k)=0\,.
\eeq
We can now inverse Fourier-transform this equation. We realise that multiplication with $\cosh \frac{\pi \omega}{2} $ in the first term is equivalent to shifting the argument of the inverse Fourier transformed function so we get
\beq
\xi(\theta+\frac{i\pi}{2})+\xi(\theta-\frac{i\pi}{2})+2L(\theta)=0\,.
\label{7}
\eeq 
Since $\cosh(\theta+\frac{i\pi}{2})+\cosh(\theta-\frac{i\pi}{2})=0$ the $m\beta \cosh\theta$ part of the $\xi(\theta)$ function is cancelled in the combination above so the equation becomes
\beq
\varepsilon(\theta+\frac{i\pi}{2})+\varepsilon(\theta-\frac{i\pi}{2})+2L(\theta)=0\,,
\eeq 
thus we now have a new TBA equation where the driving term $m\beta \cosh\theta$ does not appear explicitly. Instead, we can rewrite the equation is a more elegant form by defining the function $Y(\theta)=e^{\varepsilon(\theta)}$ which then gives
\beq 
Y(\theta+\frac{i\pi}{2})Y(\theta-\frac{i\pi}{2})(1+\frac{1}{Y(\theta)})^2=1\,.
\eeq 
Information about the driving term is now encoded in the asymptotics of the function $Y(\theta)$. In addition, it is known from multiple works, including \cite{tba1,elastic,tba2}, that  for high temperature $\beta \ll 1$ and strongly peaked kernels such as ours,  the function $Y(\theta)$ tends to a value $y$ which is constant \footnote{In this limit the TBA equations are known as constant TBA equations \cite{tba1,elastic,tba2}.} and so the $Y$-system becomes instead an algebraic equation for the number $y$, which for sinh-Gordon has solution $y=0$.  Famously, this solution can then be plugged into Roger's dilogarithm to produce the central charge of the ultraviolet fixed point \cite{tba1,elastic,tba2}, which for sinh-Gordon is that of a massless free boson $c=1$.

An interesting observation about this derivation is that it follows through in exactly the same way if we replace the driving term $m\beta \cosh\theta$ by
\beq 
\nu(\theta)=\sum_{s \, {\rm odd}} m^{s}\left( \lambda^+_s e^{s\theta} + \lambda_s^- e^{-s\theta} \right)\,,
\eeq 
since, $\exp(\pm s(\theta+\frac{i\pi}{2}))+\exp(\pm s(\theta-\frac{i\pi}{2}))=0$ for $s$ odd. The constants $\lambda^\pm_s$ are such that the combination $m^s \lambda^\pm_s$ is dimensionless. Note that the sum above represents a sum over any subset of odd spin eigenvalues, so it represents an infinite family of distinct driving terms $\nu(\theta)$. Thus, a GGE involving any combination of odd spin charges, which for this model are known to be the local conserved charges, gives rise to the same $Y$-system as in a thermal state. Equivalently, any GGE involving any spin $s$ conserved charges where $s$ is not odd (that is, non-local conserved charges in the present example), will give rise via the procedure above, to a ``modified" $Y$-system, where the driving terms of the original TBA equations would not be fully cancelled out. This is similar to the situation described in \cite{IQC}, particularly equation (14) therein. 

\medskip
Given the prevalence of $Y$-functions in the context of quantum spin chains a small clarification is due. Unlike for quantum spin chains where the Bethe ansatz equations are frequently written in terms of $Y$-functions even before any Fourier transformation has been applied, in IQFTs, following Zamolodchikov's original proposal \cite{Z} the term $Y$-system is only employed to refer to the equations that result from expressing the TBA equations in a form where there are no longer convolution terms and where the equations depend only on $Y$-functions and their shifted versions. 

\medskip 
The aim of this letter is to highlight the fact that this phenomenon, whereby infinitely many different GGE's are described by the same $Y$-system is by no means restricted to the sinh-Gordon model, but can be found for many IQFTs.  The fact that more generic TBA driving terms would also cancel out in the $Y$-system construction is not knew, in the sense that the required cancellation properties for such terms were noticed on at least some early TBA papers. However, since the idea of considering more general driving terms was not entertained at the time, the observation was never set in the context of GGEs.  We will take as our main example the family of minimal Toda field theories. 

\section{Toda Field Theory}
In this section we present a brief history of massive Toda field theories and their $S$-matrix construction and then review the $Y$-systems of minimal Toda field theories (MTFTs) of the ADE family. Finally, we present a simple example. 
\subsection{A Bit of History}
Most IQFTs and conformal field theories can be related in some way to the Toda family. For massive theories with diagonal scattering, there is generally a connection to either MTFTs or affine Toda field theories. The most famous and interesting feature of Toda theories is their underlying algebraic structure. This structure allows for establishing a one-to-one relationship between physical quantities/properties and Lie-algebraic quantities/properties. The simplest examples are the Ising model, which can be seen as the simplest MTFT, associated with the algebra $a_1$ and the sinh-Gordon model, which is the simplest example of an affine Toda field theory, also associated with the algebra $a_1$\footnote{At the $S$-matrix level, affine Toda field theories have the same $S$-matrices as MTFTs times CDD factors which incorporate a dependence on the coupling but no additional physical poles. Thus the sinh-Gordon $S$-matrix can be seen as a pure CDD factor times the (trivial) Ising $S$-matrix. The self-dual point mentioned earlier corresponds to fixing the coupling constant to the special value $B=1$ (under appropriate normalisation).}. 

Both in MTFT and affine Toda field theories the particle spectrum, $S$-matrices and conserved charges are in one-to-one correspondence with algebraic data. A sketch of this correspondence is given below:

\begin{table}[h!]
    \centering
    \begin{tabular}{|ccc|}
    \hline
    &&\\
      \# Nodes in the Dynkin diagram    & 
      $\Longleftrightarrow$ & \# of particles in the spectrum \\
      &&\\
      \hline\hline 
         &&\\
      Symmetry of the Dynkin diagram     & $\Longleftrightarrow$ & particle charge conjugation \\
      &&\\
      \hline\hline 
      &&\\
      Cartan and incidence matrices   & $\Longleftrightarrow$ & universal $S$-matrix structure \\
      &&\\
      \hline\hline 
      &&\\
      Coxeter number $h$   & $\Longleftrightarrow$ & $S$-matrix poles at multiples of $\theta=\frac{i\pi}{h}$ \\
      &&\\
      \hline 
    \end{tabular}
\end{table}
A family of extensively studied MTFTs are those associated with the simply-laced finite algebras $a_n, d_n, e_6, e_7$ and $e_8$ ($n \in \mathbb{Z}^+$). These are the ADE scattering theories  whose TBA was studied in \cite{elastic, Z,RTV}. 

The construction of the exact scattering matrices of these theories has a long history as different models/algebras were understood at different times \cite{KoSw,start,potts,FKM, FaZa,ChMu,ChMu1,BCDS,elastic}, and later brought under a more general construction \cite{Do1,FLO,Do2,FO}. Still later a ``universal" scattering matrix was proposed which encoded within a single algebraic representation the $S$-matrices of all affine Toda field theories related to both simply- and non-simply laced algebras  \cite{FKS} and the corresponding TBA was studied in \cite{FKS2}.

\subsection{Y-Systems for ADE Theories}
The scattering matrices of ADE minimal Toda field theories can be found in many papers, as discussed above. They take the universal form 
\beq
S_{ab}(\theta)= \exp\left[\int_{-\infty}^\infty \frac{dk}{k} 2 \cosh\frac{\pi k}{h} \left(2\cosh\frac{\pi k}{h}-{\bf I} \right)^{-1}_{ab} e^{-ik \theta}\right] \,,
\eeq
where ${\bf I}$ is the incidence matrix of the associated algebra\footnote{Not to be mistaken for the identity matrix!}, which in turn relates to the connectivity of its Dynkin diagram, and $h$ its Coxeter number. This representation is particularly useful in order to obtain the scattering phases and, especially, their Fourier transform, which follows directly as
\beq
\tilde{\varphi}_{ab}(k)=\delta_{ab}-2 \cosh\frac{\pi k}{h} \left(2\cosh\frac{\pi k}{h}-{\bf I} \right)^{-1}_{ab}\,.
\eeq 
Starting from this $S$-matrix it has been shown in \cite{Z,RTV} that for the case of a TBA-equation of the form (\ref{tba}),  with driving term $\nu_a(\theta)=m_a \cosh\theta$, the corresponding $Y$-system is, 
\beq
Y_a(\theta+\frac{i\pi}{h}) Y_a(\theta-\frac{i\pi}{h})=\prod_{b=1}^r (1+Y_b(\theta))^{{\bf I}_{ab}}\,,
\label{ysys}
\eeq
The particular simplification that results from Fourier-transforming and then inverse Fourier-transforming the TBA equations while, in the process, getting rid of the convolution integral, is more involved in this case. Once more, the end result is that the driving term does not feature any longer. The cancellation reported after (\ref{7}) is now replaced by the equation
\beq
m_a \cosh(\theta+\frac{i\pi}{h})+m_a \cosh( \theta-\frac{i\pi}{h})- \cosh\theta\sum_{b=1}^r {\bf I}_{ab} m_b =0\,,
\eeq
where we used the famous identity
\beq
\sum_{b=1}^r {\bf I}_{ab} m_b =2m_a \cos\frac{\pi}{h}\,.
\label{15}
\eeq
This identity is equivalent to the statement that the masses of the theory form an eigenvector of the incidence matrix ${\bf I}$ with eigenvalue $2\cos\frac{\pi}{h}$.  This is the Perron-Frobenius eigenvector\footnote{The Perron-Frobenius theorem states that a real square matrix with positive entries has a unique largest real eigenvalue and that the corresponding eigenvector can be chosen to have strictly positive components.} of the incidence matrix and is also an eigenvector of the Cartan matrix of the corresponding algebra, which is related to ${\bf I}$ by ${\bf C}:=2-{\bf I}$.

An observation that can be found explicitly in \cite{RTV} is that the other eigenvectors of the incidence/Cartan matrix can be also constructed in terms of the higher spin conserved quantities of the theory. More precisely, let $q_a^s$ be the one-particle eigenvalue for particle $a$ of a conserved charge of spin $s$ at rapidity zero. Then, 
\beq
\sum_{b=1}^r {\bf I}_{ab} q_b^s= 2 q_a^s \cos\frac{\pi s}{h}\,.
\label{highers}
\eeq 
Although this has been known for a long time, it has not been connected explicitly to GGEs. Let us now consider the same TBA-equations with driving term
\beq
\nu_a(\theta)=\sum_{s\,{\rm odd}} \lambda_s q_a^s  \cosh(s \theta)\,,
\label{driving}
\eeq
where the sum is meant over any subset of values of $s$ as long as they are odd.
Then, the resulting $Y$-system is still (\ref{ysys}) and in this case, the dependence on the driving term drops out thanks to the identity
\beq
q_a^s \cosh(s\theta+\frac{i\pi s}{h})+q_a^s \cosh( s\theta-\frac{i\pi s}{h})- \cosh(s\theta)\sum_{b=1}^r {\bf I}_{ab} q_b^s=0\,,
\label{qs}
\eeq 
which holds thanks to (\ref{highers}) for odd spin. Note that the same conclusion holds if we also allow $\sinh$-terms in (\ref{driving}), as long as the spin is odd. Including such terms is equivalent to allowing for Lorenz boosts of the original state.

The numbers $q_a^s$ have been systematically computed (or can be constructed from other results), for instance in \cite{BCDS}. Therefore, the conclusion is that there is an infinite family of GGEs, characterised by driving terms involving {\it any number} of appropriately normalised one-particle eigenvalues of odd local conserved charges, which gives rise to the same $Y$-system, in the sense of satisfying (\ref{ysys}). It is worth noting however that the analyticity properties and meaningfulness of the $Y$-system need to be considered carefully when the number of terms in (\ref{driving}) is infinite. For quantum spin chains, the important subtleties arising in this situation have been highlighted in \cite{IQC,LBE}. Of course, the $Y$-system is characterised not only by the equation (\ref{ysys}) but by the asymptotics of the functions $Y_a(\theta)$ which depends on the specific driving term. However, there are fundamental properties of the $Y$-system which only depend on the equation above, such as its periodicity and asymptotics for small Lagrange multipliers $\lambda_s$.

\medskip

Although we have restricted ourselves to ADE MTFTs here, the same type of arguments can be made for affine Toda field theories, which in fact share the same property (\ref{qs}) and for other theories which share the algebraic structure of Toda models. Examples are the homogeneous sine-Gordon models \cite{Smatrix, ourtba,ourtba2,DoMi} generalisations thereof, such as the ``colour-valued" $S$-matrices proposed in \cite{CoVa,CoVa2} and other Dynkin diagram based theories, some of which are also discussed in \cite{RTV}. Our conclusions also extend to integral equations of the type found for the sine-Gordon/XXZ models \cite{DdV1,Klu,DdV2}. 

\subsection{$a_{r}$ Minimal Toda Field Theory}
It is interesting to report the basic data we were talking about in the previous subsection for one particular family of theories. For $a_{r}$-MTFTs the Cartan and incidence matrices are:
\beq 
{\bf C}_{ab}=2\delta_{ab}-{\bf I}_{ab}=2\delta_{ab}-\delta_{a\, b+1}-\delta_{a \,b-1}\,, \qquad a,b=1,\ldots,r\,,
\eeq
and the mass spectrum and higher conserved charges (at zero rapidity) are given by:
\beq 
m_a=2m\sin\frac{\pi a}{r+1}\qquad {\rm and} \qquad q_a^s=2 m^s{\sin\frac{\pi a s}{r+1}}\,,
\label{this}
\eeq 
where $m$ is a mass scale. The Coxeter number of the algebra is $h=r+1$ in terms of the rank $r$.
The formulae
(\ref{this}) can be easily shown to satisfy the equations (\ref{15}) and (\ref{highers}), respectively. Note that equation (\ref{highers}) is satisfied for all integer values of $s$, not only odd ones. 

\section{Conclusions and Outlook}

In this paper we have pointed out a mathematical curiosity regarding the generalisation of TBA equations to GGEs in IQFTs. Our observation is that an infinite family of different TBA equations gives rise to the same $Y$-system as for a Gibbs ensemble. Interestingly, this family selects out for TBA driving terms which involve only the one-particle eigenvalues of local, odd charges, that is, it corresponds to GGEs involving only odd spin local charges. Thus, in a certain sense, the $Y$-systems of IQFTs select for the local conservation laws of a particular theory. 

For each TBA equation, the corresponding $Y$-functions satisfy asymptotic properties which are dependent on which Lagrange multipliers are large/small, therefore we can also think of this result as a way of counting the number of distinct solutions to a particular $Y$-system. The answer seems to be that there are as many independent solutions as there are subsets of odd local conserved charges in the model. Although the derivation presented here does not particularly discuss the case of an infinite number of odd spin conserved changes, there are known subtleties that must be carefully considered in this situation, as noted in \cite{IQC,LBE}.

This result is interesting because it is known that all solutions to the same $Y$-system share some properties. As shown in \cite{Z} for ADE MTFTs the solutions of (\ref{ysys}) are periodic in $\theta$, with periodicity which is proportional to the conformal dimension of the perturbing field. It is also the case that in any limit where the $Y$-functions tend to a constant (which would correspond to all generalised temperatures being large), the solutions to the constant $Y$-system are the input needed to compute the UV central charge of the model. 

In the context of IQFT, $Y$-systems do not seem to be particularly useful in order to solve the TBA equations, so our result is probably of little value from a computational viewpoint. However it does point to a common algebraic structure that is shared by large class of GGEs involving only odd local charges and this is intriguing. It would be interesting to understand if and how this structure can tell us something new about the (shared) mathematical and physical properties of these types of stationary states. In particular, we might ask whether they are related to other distinct families of GGEs whose physical interpretation is clearer, such as those generated by so-called integrable quenches \cite{LBE2}. This relationship is by no means obvious at this stage, especially because the analysis of \cite{LBE2} is focussed on lattice models, but would be interesting to investigate for IQFTs. 

\medskip

\noindent {\bf Acknowledgment:} I am grateful to Roberto Tateo for a very useful discussion at the early stages of this project, to Bal\'azs Pozsgay for bringing to my attention the subtleties of considering GGEs involving infinitely many conserved charges and to Jacopo De Nardis for suggesting that the special class of GGEs found here could be related to the integrable quenches introduced in \cite{LBE2}. I thank Jacopo De Nardis, Benjamin Doyon, Michele Mazzoni and Takato Yoshimura for discussions and interest in this work. Some of these discussions took place at the Galileo Galilei Institute in Florence during the programme on Randomness, Integrability and Universality, April 19--June 3 (2022). I am indebted to the organisers of this programme for inviting me to participate and for financially supporting my stay. Finally, I am grateful to Bal\'azs Pozsgay and Lorenzo Piroli for bringing several relevant references to my attention. 


\end{document}